\documentstyle[aasms4,12pt]{article}
\begin{document}

\title{OPTICAL VERSUS MID-INFRARED SPECTROSCOPIC CLASSIFICATION OF 
       ULTRALUMINOUS INFRARED GALAXIES} 

\author{\sc Yoshiaki Taniguchi, Akira Yoshino, Youichi Ohyama, \& Shingo Nishiura}

\affil{Astronomical Institute, Tohoku University, Aoba, Sendai 980-8578, Japan}


\begin{abstract}
The origin of huge infrared luminosities of ultraluminous infrared
galaxies (ULIGs) is still in question. 
Recently, Genzel et al. made mid-infrared (MIR) spectroscopy of 
a large number of ULIGs and found that the major energy source in them
is massive stars formed in the recent starburst activity;
i.e., $\sim$ 70\% -- 80\% of the sample are predominantly powered by the
starburst. However, it is known that previous optical spectroscopic 
observations showed that the majority of ULIGs are classified as Seyferts
or LINERs (low-ionization nuclear emission-line regions).
In order to reconcile this difference,
we compare types of emission-line activity for a sample of ULIGs which have 
been observed in both optical and MIR.
We confirm the results of previous studies that the majority of ULIGs
classified as LINERs based on the optical emission-line 
diagnostics turn to be starburst-dominated galaxies based on the MIR ones. 
Since the MIR spectroscopy can probe more heavily-reddened, inner parts of 
the ULIGs, it is quite unlikely that the inner parts are powered by the
starburst while the outer parts are powered by non-stellar ionization sources.
The most probable  resolution of this dilemma is that the optical 
emission-line nebulae  with the LINER properties are powered predominantly by
shock heating driven by the superwind activity;
i.e., a blast wave driven by a collective effect of a large number of supernovae
in the central region of galaxy mergers.
\end{abstract}


\keywords{
galaxies: emission lines {\em -}
galaxies: starburst {\em -} galaxies: active {\em -} galaxies: Seyfert {\em -}
infrared {\em -} emission}


\section{INTRODUCTION}

The origin of huge infrared luminosities of ultraluminous infrared
galaxies (ULIGs) is still in question; i.e., which is the major
energy source, starbursts or active galactic nuclei (AGN) ?
(Sanders et al. 1988a; Condon et al. 1991; Majewski et al. 1993;
Smith et al. 1998; see for a review Sanders \& Mirabel 1996).
Since their far-infrared luminosities, $L$(FIR) $\gtrsim 10^{12} L_\odot$,
are comparable to those of quasars, possible starburst-AGN connections
have also been discussed by many authors (Sanders et al. 1988a, 1988b;
Norman \& Scoville 1988;
Terlevich et al. 1992; Lonsdale, Smith, \& Lonsdale 1993; Lonsdale et al. 1998;
see, however, also Section 4 in Taniguchi 1997).

Optical spectroscopy has been often used to investigate the activity in 
emission-line galaxies (e.g., Veilleux \& Osterbrock 1987).
In fact, optical spectroscopic observations have shown that the majority of 
the ULIGs are AGN-like objects; Seyferts or LINERs\footnote{LINER = 
Low Ionization Nuclear Emission-line Region (Heckman 1980).} (Sanders et al. 1988a;
Armus et al. 1989; Kim et al. 1995; Kim, Veilleux, \& Sanders 1998;
Veilleux et al. 1995; Veilleux 1997).
Further, in order to find hidden broad emission line regions (BLRs), which provide
definite evidence for AGN, near-infrared (NIR) spectroscopic programs have
also been conducted (Goldader et al. 1995, 1997a, 1997b; 
Veilleux, Sanders, \& Kim 1997). 
However, hidden BLRs have  been found in several ULIGs.
Therefore, the majority of AGN-like ULIGs are type 2 Seyferts (S2s) or LINERs
although the fraction of type 1 Seyferts (S1s) increases with increasing 
$L$(FIR) (Kim et al. 1998).
ULIGs which are optically classified as S1s or S2s are not necessarily
powered by an AGN.  Their classification only indicates the existence of an AGN.
This AGN may not dominate the total bolometric luminosity.

Furthermore, since some LINER-like ULIGs show the extranuclear LINER emission
(Veilleux et al. 1995; Kim et al. 1998; Heckman et al. 1996; Taniguchi \& Ohyama 1998),
the origin of LINER-like ULIGs is controversial because 
this emission can be attributed to shocks caused by the interaction between
starburst-driven outflows and the ambient gas. 
It is also observed that a number of ULIGs have intense star forming regions 
in their central regions (e.g., Shaya et al. 1994; Surace et al. 1998).
For example, the main energy source in Arp 220, one of the archetypical ULIGs, 
is attributed to the intense starburst in the central region
(Shaya et al. 1994; Larkin et al. 1995; Skinner et al. 1997; Iwasawa 1998;
Scoville et al. 1998; Taniguchi, Trentham, \& Shioya 1998;
Shioya, Taniguchi, \& Trentham 1998). However, possible evidence for 
the hidden AGN in Arp 220 has also been discussed 
because of the presence of compact OH megamaser sources in its central region
(Diamond et al. 1989; Lonsdale et al. 1994, 1998). 
This raises another important question;
what powers the ULIGs on average ? (Genzel et al. 1998).
Although this is actually an important question, it is difficult to give
a certain answer because all the ULIGs have a large amount of 
molecular gas and dust in their central regions (Scoville et al. 1991;
Scoville, Yun, \& Bryant 1997;
Downes \& Solomon 1998 and references therein) and thus the putative 
central engine could be obscured even if present.
Therefore, if we would like to probe the very inner regions of ULIGs,
we have to perform MIR and FIR spectroscopy of ULIGs.

Recently, Genzel et al. (1998) made MIR spectroscopy of 
a large number of ULIGs. Since the MIR spectroscopy can probe heavily
reddened ($A_V \sim$ 50 mag) regions, it would be very powerful in searching
for hidden AGN in ULIGs (see also Lutz et al. 1996). 
However, they found that their major energy source
is massive stars formed in the recent starburst activity;
i.e., $\sim$ 70\% -- 80\% of the sample are predominantly powered by the
starburst. On the other hand, as described before, the previous optical spectroscopic 
observations showed that the majority of ULIGs are classified as Seyferts
or LINERs. In order to reconcile this difference,
we compare types of emission-line activity for a sample of ULIGs which have 
been observed in both optical and MIR and 
discuss a possible resolution of this dilemma in this paper.


\section{A BRIEF SUMMARY OF THE MIR SPECTROSCOPY BY Genzel et al. (1998)}

Genzel et al. (1998) performed the MIR spectroscopic survey of 15 ULIGs using 
both short wavelength grating spectrometer (SWS) and ISOPHOT-S spectrometer
on board Infrared Space Observatory (ISO). They utilized either 
[Ne {\sc v}]$\lambda$14.3$\mu$m/[Ne {\sc ii}]$\lambda$12.8$\mu$m or 
[O {\sc iv}]$\lambda$25.9$\mu$m/[Ne {\sc ii}]$\lambda$12.8$\mu$m 
emission-line intensity ratio as the discriminator between starbursts 
and AGN\footnote{They 
note that the [S {\sc iii}]$\lambda$33.5$\mu$m line can also be used
instead of [Ne {\sc ii}]$\lambda$12.8$\mu$m whenever the [Ne {\sc ii}] 
line data are not available because the [S {\sc iii}] emission arises
nearly the same ionization-condition region as that of the [Ne {\sc ii}] 
emission.}.
Since the poverty of high energy photons  from massive OB stars,
high-ionization lines such as [Ne {\sc v}] are generally weak 
in starburst galaxies. On the other hand, such high-ionization lines
are often strong in AGN because of a larger number of high-ionization
photons (Spinoglio \& Malkan 1992; Voit 1992a).
In fact, Genzel et al. (1998) obtained
[Ne {\sc v}]$\lambda$14.3$\mu$m/[Ne {\sc ii}]$\lambda$12.8$\mu$m 
$<$ 0.01 for a sample of nearby typical starburst galaxies (e.g., M82)
while $\sim$ 0.1 -- 1 for a sample of nearby AGN (e.g., NGC 4151).

They detected the [Ne {\sc v}] emission only from Mrk 273 
among the 15 ULIGs (see Table 1). Its [Ne {\sc v}]/[Ne {\sc ii}] ratio, 0.27,
is consistent with the observed range for AGN. 
Although the [O {\sc iv}] was detected in IRAS 23128$-$5919,
its [O {\sc iv}]/[Ne {\sc ii}] ratio, 0.037, is so small that 
this ULIG is a starburst-dominated galaxy.
Upper limits for the remaining objects lie in the intermediate 
range between the starbursts and AGN, indicating the intermediate
nature of the ULIGs.
In order to estimate the relative contribution of starbursts
to the total energy, they also used  the 7.7 $\mu$m emission feature which is
one of unidentified IR emission bands, probably emitted by
polycyclic aromatic hydrocarbons (PAHs). Since these features
can be seen only in star-forming galaxies, they are useful
in estimating the star formation activity in galaxies (Moorwood \& Oliva 1988;
Mouri et al. 1990; Voit 1992b).
Then using the diagram of the [O {\sc iv}]/[Ne {\sc ii}] intensity ratio
vs. the relative strength of 7.7 $\mu$m PAH feature, 
they concluded that 70\% -- 80\% 
of their ULIG sample are dominated by the starburst component. 

\section{OPTICAL vs. MIR CLASSIFICATIONS OF ULIGs}

We compare the optical emission-line classification with 
the MIR one for a sample of ULIGs which are observed in both optical and MIR.
Our sample consists of 12 ULIGs given in Table 1.
We have compiled the following important MIR and optical emission-line
ratios; a) MIR: 1) [Ne {\sc v}]$\lambda$14.3$\mu$m/[Ne {\sc ii}]$\lambda$12.8$\mu$m,
and 2) [O {\sc iv}]$\lambda$25.9$\mu$m/[Ne {\sc ii}]$\lambda$12.8$\mu$m
(Genzel et al. 1998), and b) optical: 1) [O {\sc iii}]$\lambda$5007/H$\beta$,
2) [N {\sc ii}]$\lambda$6583/H$\alpha$, 
3) [S {\sc ii}]$\lambda\lambda$6717,6731/H$\alpha$, 
and 4) [O {\sc i}]$\lambda$6300/H$\alpha$ (Sanders et al. 1988a; Armus et al. 1989;
Veilleux et al. 1995; Duc et al. 1997; Kim et al. 1998). 
As shown in Table 1, since the most MIR line ratios are upper-limit data,
it is difficult to classify the ULIGs unambiguously solely with the ratios.
Therefore, following Genzel et al. (1998), we use the diagram 
between the [O {\sc iv}]/[Ne {\sc ii}] intensity ratio
and the relative strength of 7.7 $\mu$m PAH feature (see Figure 5 in
Genzel et al. 1998). This diagram shows that  only two ULIGs, Mrk 231 and Mrk 273, 
are classified as AGN; i.e., higher than 50\% of the energy comes from AGN.
The optical classification is made with  
the criteria given in Figure 5 in Veilleux et al. (1995;
see also Veilleux \& Osterbrock 1987).

The optical line ratios show that 
only four galaxies are classified as H{\sc ii}-region (i.e., starbursts)
galaxies. Mrk 231\footnote{The optical spectrum
of Mrk 231 is dominated by very strong Fe {\sc ii} emission features, which
are one of important characteristics of type 1 AGN (e.g., Boroson \& Green 1992).
Further, the presence of the broad Balmer emission lines provides the evidence for 
AGN in this galaxy (Boksenberg et al. 1977; L\'ipari, Colina, \& Macchetto 1994
and references therein).} and Mrk 273 are S1 and S2, respectively.
The remaining six ULIGs are LINERs. 
In Table 2, we compare the activity types based 
on the MIR classification with those based on the optical one. 
Both the MIR and optical classification give the same activity types
for the following six galaxies; Mrk 231 and Mrk 273 (AGN), and 
IRAS 17208$-$0014, IRAS 20100$-$4156, IRAS 20551$-$4250, and IRAS 22491$-$1808 
(starburst) although IRAS 20551$-$4250 has been suspected to have LINER-like 
properties (Johansson 1991; Duc et al. 1997).
Since the MIR observations can probe the heavily-reddened, inner part of 
the nuclear region, we expect that the MIR spectroscopy would reveal 
hidden AGN in some ULIGs which are classified as starburst galaxies based
on the optical observations. However, no such case is found.
Contrary to the above expectation, six LINER-like ULIGs based on the optical
observations turn to be starburst-dominated galaxies based on the MIR spectroscopy.
This suggests strongly that these ULIGs which are optically classified as LINERs 
are not genuine AGN but shock-heated galaxies. If this is the case, we can conclude 
that both the MIR and optical classifications give the same activity types 
for all the ULIGs analyzed in this study.

We mention the difference between the MIR and optical classification
schemes. Since optical emission lines arise mostly from low-ionization ions,
the optical classification uses the low-ionization emission-line ratios
(Veilleux \& Osterbrock 1987; Veilleux et al. 1995).
On the other hand, Genzel et al. (1998) adopted the high-ionization lines
such as [O {\sc iv}] and [Ne {\sc v}] in order to obtain firmer evidence for
the central engine of AGN.
Therefore, if we would like to identify LINERs in future MIR spectroscopy,
we will have to choose some low-ionization MIR emission lines such as
[O {\sc i}]$\lambda 63\mu$m, [O {\sc i}]$\lambda 145\mu$m, and
[C {\sc ii}]$\lambda 158\mu$m (Spinoglio \& Malkan 1992; Luhman et al. 1998).

\section{DISCUSSION}

The most important implication suggested by the MIR spectroscopy is that 
the majority of ULIGs have little evidence for AGN with high-ionization 
emission lines.
This casts a serious problem;  ^^ ^^ Are low-ionization AGN classified
with the optical spectroscopy really AGN ?"
Since the discovery of LINERs (Heckman 1980), this question has been
addressed many times (Heckman 1986; Ho 1998 and references therein).
Here we consider what the case for the ULIGs is\footnote{We do not discuss about
LINERs associated with nuclei of ordinary galaxies in this paper. 
Those who are interested in this
issue are recommended to see Ho, Filippenko, \& Sargent (1997),
Maoz et al. (1998), and  Barth et al. (1998).}.
There are three alternative ideas to explain the LINER-like spectra 
in the optical; a) a genuine AGN with low-ionization conditions
(e.g., Ferland \& Netzer 1983; Halpern \& Steiner 1983), b)  dense gas clouds 
photoionized by hot O stars ($T_{\rm eff} \gtrsim$ 45,000 K)
(Filippenko \& Terlevich 1992; Shields 1992), and c) shock heating
(Johansson \& Bergvall 1985, 1988; Johansson 1991; Veilleux et al. 1995, 1997;
Kim et al. 1998; see also Heckman 1980; Dopita \& Sutherland 1995).
Since the MIR classification made by Genzel et al. (1998) deals only with 
high-ionization lines, it is impossible to probe genuine LINERs
in the ULIGs. However, the most important point clarified by Genzel et al. (1998)
is that the major energy source in the ULIGs is massive stars. 
This suggests strongly that the optical emission-line nebulae
cannot be attributed mainly to the photoionization by the central engine of AGN.
Therefore, we discuss the latter two possibilities.

The intense starbursts are actually occurring in the heart of ULIGs
(Armus et al. 1989; Shaya et al. 1994; Scoville et al. 1998; Taniguchi et al. 1998).
Further, the gas densities in them are generally higher than those in
typical star forming regions in galaxies (Taniguchi \& Shioya 1998). 
Therefore, we discuss the second possibility at first. 
Although LINERs are often regarded as low-luminosity extensions of
typical luminous AGN, dense gas clouds photoionized by hot O stars
($T_{\rm eff} \gtrsim$ 45,000 K) also show LINER-like optical spectra
(Filippenko \& Terlevich 1992; Shields 1992). One of the important 
characteristics of such O-star LINERs is that the [O {\sc i}]$\lambda$6300
emission is significantly weak with respect to the H$\alpha$ emission;
[O {\sc i }]/H$\alpha < 1/6$. We mention that all the ULIGs except NGC 6240
satisfy this condition (see Table 1). Although there is no firm evidence
for hot O stars in the ULIGs, the presence of Wolf-Rayet stars is reported
for some luminous infrared galaxies; e.g., IRAS 01003-2238
(Armus, Heckman, \& Miley 1988) and NGC 1614 (Conti 1991). 
If these high-temperature descendents of
massive stars dominate the photoionization of ULIGs, they would lead to the
LINER-like spectra. In Figure 1, we compare the optical emission-line ratios
with the hot O-star photoionization models of both Filippenko \& Terlevich (1992)
and Shields (1992). The comparison shows that both the models cannot explain 
the data points of the LINER-like ULIGs. Although these models may explain the 
excitation properties of some starburst-dominated ULIGs, we conclude that 
hot O stars are  not the major ionization source in the LINER-like ULIGs studied here.

We consider the possibility of shock heating.
It is naturally expected that some ULIGs are dominated 
by shock heating because they show evidence for the superwind activity; i.e.,
a blast wave driven by a collective effect of a large number of supernovae
in the central region of galaxy mergers (Heckman, Armus, \& Miley 1987, 1990;
Veilleux et al. 1995; Heckman et al. 1996; Ohyama, Taniguchi, \& Terlevich 1997;
Kim et al. 1998).
Further, very recently, Lutz et al. (1998) detected faint 
[O {\sc iv}]$\lambda 25.9\mu$m emission in a number of nearby starburst
galaxies such as M82. This provides strong evidence for shock heating
in the starburst galaxies  because they showed that
the fast shock models by Dopita \& Sutherland (1995) explain
both the optical and MIR spectroscopic properties of M82 consistently.
In order to examine if the shock heating models by Dopita \& Sutherland (1995) 
can explain the optical emission-line properties of the ULIGs studied here,
we compare the observed optical emission-line ratios with results of
their shock models in Figure 1.
It is shown that the majority of the ULIGs can be explained by the 
shock heating models with shock velocity between 100 km s$^{-1}$ and 
500 km s$^{-1}$ (see also Taniguchi \& Ohyama 1998).

In summary, comparing the MIR and optical classifications for the sample of 
ULIGs, we have confirmed that the LINER-like ULIGs are not genuine AGN
but shock-heated galaxies as suggested by Veilleux et al. (1995) and
Kim et al. (1998). The recommendation should be to use both the optical and
MIR classifications rather than one method. 
Finally, we should mention that our discussion is applicable only to infrared 
selected emission-line galaxies. Optically-selected LINERs may be powered by
another energy source, e.g., AGN (see Ho 1998 and references therein). 

\vspace{1ex}

We would like to thank the referee, Sylvain Veilleux, for useful comments 
which improved this paper significantly. 
YT also thanks many people at the Ringberg Castle ULIG meeting 
for their many useful inputs. In particular, he thanks Reinhard Genzel,
Dieter Lutz, and Linda Tacconi for their very nice organization of the
Ringberg meeting. 
This work was supported in part by the Ministry of Education, Science,
Sports and Culture in Japan under Grant Nos. 07055044, 10044052, and 10304013.
A part of this work was made when YT visited the Astronomical Data Analysis Center 
(ADAC) of the National Astronomical Observatory of Japan. YT thanks the staff of ADAC
for their kind hospitality.

\newpage

\figcaption{
Excitation diagrams for the ULIGs except Mrk 231;
{\it a)} [N {\sc ii}]/H$\alpha$ (left), 
{\it b)} [S {\sc ii}]($\lambda$6716+6731)/H$\alpha$
(middle), and {\it c)} [O {\sc i}]$\lambda$6300/H$\alpha$ (right) are plotted
against [O {\sc iii}]$\lambda$5007/H$\beta$.
Dashed curve shows the distinction among H {\sc ii} region-like objects,
Seyfert 2 (S2), and LINERs in each panel, taken from Veilleux et al. (1995).
The alphabets correspond to those in Table 1.
The hot O-star photoionization models by Filippenko \& Terlevich (1992)
are shown by dash-double dotted curves; for the solar abundances (thin) and for 
the 1.4 times solar abundances except nitrogen which is 4 times solar (thick).
A logarithmic value of the ionization parameter is labeled at each asterisk mark.
The composite hot O-star model by Shields (1992) is shown by a open 
square in each panel.
The shock models are taken from
Dopita \& Sutherland (1995); 1) ^^ ^^ shock + precursor" models (solid
curves) with magnetic parameter $B/n^{1/2}$ = 0, 2, and 4 $\mu$G cm$^{3/2}$
and shock velocity from 200 km s$^{-1}$ to 500 km s$^{-1}$, and
2)  ^^ ^^ shock only" models (dash-dot
curves) with magnetic parameter $B/n^{1/2}$ = 0, 2, and 4 $\mu$G cm$^{3/2}$
and shock velocity from 150 km s$^{-1}$ to 500 km s$^{-1}$.
\label{fig1}
}


\newpage

\begin{references}
\reference{1}{Armus, L., Heckman, T. M., \& Miley, G. K. 1988, ApJ, 326, L45}
\reference{1}{Armus, L., Heckman, T. M., \& Miley, G. K. 1989, \apj, 347, 727}
\reference{1}{Barth, A. J., Ho, L. C., Filippenko, A. V., \& Sargent, W. L. W. 
              1998, ApJ, 496, 133}
\reference{1}{Boksenberg, A., Carswell, R., Allen, D., Fosbury, R., Penston, M.,
              \& Sargent, W. 1977, MNRAS, 178, 451}
\reference{1}{Boroson, T. A., \& Green, R. F. 1992, ApJS, 80, 109} 
\reference{1}{Condon, J. J., Huang, Z. -P., Yin, Q. F., \& Thuan, T. X.
              1991, ApJ, 378, 65}
\reference{1}{Conti, P. S. 1991, ApJ, 377, 115}
\reference{1}{Diamond, P. J., Norris, R. P., Baan, W. A., \& Booth, R. S. 1989, 
              ApJ, 340, L49}
\reference{1}{Dopita, M. A.,  \& Sutherland, R. S.  1995, \apj, 455, 468}
\reference{1}{Downes, D., \& Solomon, P. M. 1998, \apj, 507, 615}
\reference{1}{Duc, P.-A., Mirabel, I. F., \& Maza, J. 1997, A\&AS, 124, 533}
\reference{1}{Ferland, G. J., \& Netzer, H. 1983, ApJ, 264, 105}
\reference{1}{Filippenko, A. V., \& Terlevich, R. 1992, ApJ, 397, L79}
\reference{1}{Genzel, R., et al. 1998, ApJ, 498, 579}
\reference{1}{Goldader, J. D., Joseph, R. D., Doyon, R., \& Sanders, D. B.
              1995, ApJ, 444, 97}
\reference{1}{Goldader, J. D., Joseph, R. D., Doyon, R., \& Sanders, D. B.
              1997a, ApJ, 474, 104}
\reference{1}{Goldader, J. D., Joseph, R. D., Doyon, R., \& Sanders, D. B.
              1997b, ApJS, 108, 449}
\reference{1}{Halpern, J. P., \& Steiner, J. E. 1983, ApJ, 269, 37}
\reference{1}{Heckman, T. M. 1980, A\&A, 87, 152}
\reference{1}{Heckman, T. M. 1986, PASP, 98, 159}
\reference{1}{Heckman, T. M., Armus, L., \& Miley, G. K. 1987, AJ, 93, 276}
\reference{1}{Heckman, T. M., Armus, L., \& Miley, G. K. 1990, ApJS, 74, 833}
\reference{1}{Heckman, T. M., Dahlem, M., Eales, S. A., Fabbiano, G., \& Weaver, K.
              1996, ApJ, 457, 616}
\reference{1}{Ho, L. C. 1998, in Proceedings of the 23rd COSPAR Meeting,
              The AGN-Galaxy Connection (Advances in Space Research),
              in press (CfA preprint No. 4710)}
\reference{1}{Ho, L. C., Filippenko, A. V., \& Sargent, W. L. W. 1997, 
              ApJ, 487, 568}
\reference{1}{Iwasawa, K. 1998, MNRAS, in press (astro-ph/9808313)}
\reference{1}{Johansson, L. 1991, A \& A, 241, 381}
\reference{1}{Johansson, L., \& Bergvall, N. 1985, A \& A, 149, 475}
\reference{1}{Johansson, L., \& Bergvall, N. 1988, A \& A, 192, 81}
\reference{1}{Kim, D.-C., Sanders, D. B., Veilleux, S., Mazzarella, J. M., \&
              Soifer, B. T. 1995, ApJS, 98, 129}
\reference{1}{Kim, D.-C., Veilleux, S., \& Sanders, D. B. 1998, ApJ, in press 
              (astro-ph/9806149)}
\reference{1}{Larkin, J. E., Armus, L., Knop, K., Matthews, K., \& Soifer, B. T.
              1995, ApJ, 452, 599}
\reference{1}{L\'ipari, S., Colina, L., \& Macchetto, F. 1994, ApJ, 427, 174}
\reference{1}{Lonsdale, C. J., Diamond, P. J.,  Smith, H. E., \& Lonsdale, C. J. 
              1994, Nature, 370, 117}
\reference{1}{Lonsdale, C. J., Lonsdale, C. J., Diamond, P. J., \&  Smith, H. E.
              1998, ApJ, 493, L13}
\reference{1}{Lonsdale, C. J., Smith, H. E., \& Lonsdale, C. J. 1993, 
              ApJ, 405, L9}
\reference{1}{Luhman, M. L., et al. 1998, ApJ, 504, L11}
\reference{1}{Lutz, D., et al. 1996, A\&A, 315, L137}
\reference{1}{Lutz, D., Kunze, D., Spoon, H. W. W., \& Thornley, M. D.
              1998, A\&A, 333, L75}
\reference{1}{Majewski, S. R., Hereld, M., Koo, D. C., Illingworth, G. D.,
              \& Heckman, T. M. 1993, ApJ, 402, 125}
\reference{1}{Maoz, D., Koratkar, A., Shields, J. C., Ho, L. C., Filippenko, A. V.,
              \& Sternberg, A. 1998, ApJS, 116, 55}
\reference{1}{Moorwood, A. F. M., \& Oliva, E. 1988, A\&A, 203, 278}
\reference{1}{Mouri, H., Kawara, K., Taniguchi, Y., \& Nishida, M. 1990, ApJ, 356, 39}
\reference{1}{Norman, C. A., \& Scoville, N. Z. 1988, ApJ, 332, 124}
\reference{1}{Ohyama, Y., Taniguchi, Y., \& Terlevich, R. 1997, ApJ, 480, L9}
\reference{1}{Sanders, D. B., et al. 1988a, ApJ, 325, 74}
\reference{1}{Sanders, D. B., Soifer, B. T., Elias, J. H., Neugebauer, G.,
              Matthews, K. 1988b, ApJ, 328, L35}
\reference{1}{Sanders, D. B., \& Mirabel, I. F. 1996, ARA \& A, 34, 749}
\reference{1}{Scoville, N. Z., et al. 1998, \apj, 492, L107}
\reference{1}{Scoville, N. Z., Sargent, A. I., Sanders, D. B., \& Soifer, B. T.
              1991, ApJ, 366, L5}
\reference{1}{Scoville, N. Z., Yun, M. S., \& Bryant, P. M. 1997, ApJ, 484, 702}
\reference{1}{Shaya, E. J., Dowling, D. M., Currie, D. G., Faber, S. M.,
              \& Groth, E. J. 1994, AJ, 1675}
\reference{1}{Shields, J. C. 1992, ApJ, 399, L27}
\reference{1}{Shioya, Y., Taniguchi, Y., \& Trentham, N. 1998, ApJ, submitted}
\reference{1}{Skinner, C. J., Smith, H. A., Sturm, E., Barlow, M. J., 
              Cohen, R. J., \& Stacey, G. J. 1997, Nature, 386, 472}
\reference{1}{Smith, H. E., Lonsdale, C. J., \& Lonsdale, C. J. 1998, ApJ, 492, 137}
\reference{1}{Spinoglio, L., \& Malkan, M. A. 1992, ApJ, 399, 504}
\reference{1}{Surace, J. A., Sanders, D. B., Vacca, W. D., Veilleux, S., \&
              Mazzarella, J. M. 1998, ApJ, 492, 116}
\reference{1}{Taniguchi, Y. 1997, ApJ, 487, L17}
\reference{1}{Taniguchi, Y., \& Ohyama, Y. 1998, ApJ, 508, L000, in press}
\reference{1}{Taniguchi, Y., \& Shioya, Y. 1998, ApJ, 501, L167}
\reference{1}{Taniguchi, Y., Trentham, N., \& Shioya, Y. 1998, ApJ, 504, L79}
\reference{1}{Terlevich, R., Tenorio-Tagle, G., Franco, J., \& Melnick, J.
              1992, MNRAS, 255, 713}
\reference{1}{Veilleux, S. 1997, IAU Symp., 186, Galaxy
              Interactions at Low and High Redshift, ed. D. B. Sanders, \&  
              J. Barnes (Klewer: Dordrecht), 96}
\reference{1}{Veilleux, S., Kim, D.-C., Sanders, D. B., Mazzarella, J. M., \&
              Soifer, B. T. 1995, ApJS, 98, 171}
\reference{1}{Veilleux S., Osterbrock D.E., 1987, ApJS, 63, 295} 
\reference{1}{Veilleux, S., Sanders, D. B., \& Kim, D.-C. 1997, ApJ, 484, 92}
\reference{1}{Voit, G. M. 1992a, ApJ, 399, 495}
\reference{1}{Voit, G. M. 1992b, MNRAS, 258, 841}
\reference{1}{Young, S., Hough, J. H., Efstathiou, A., Wills, B. J., Bailey, J. A.,
              Ward, M. J., \& Axon, D. J. 1996, MNRAS, 281, 1206}
\end{references}
\end{document}